# Thermal enhancement of the antiferromagnetic exchange coupling between Fe epilayers separated by a crystalline ZnSe spacer

J. Varalda,[1] J. Milano,[2] A. J. A. de Oliveira,[1] E. M. Kakuno,[3] I. Mazzaro,[3] D. H. Mosca,[3] L.B. Steren,[2] M. Eddrief,[4] M. Marangolo,[4] D. Demaille,[4] V. H. Etgens [4]

[1] *Departamento de Física – UFSCar, C. P. 676, 13565-905 São Carlos SP, Brazil*

[2] *Centro Atómico Bariloche and Instituto Balseiro (CNEA-UNC), (8400) S.C. de Bariloche, Argentina*

[3] *Departamento de Física – UFPR, C. Pl. 19091, 81531-990 Curitiba PR, Brazil*

[4] *INSP- CNRS, Universités Paris VI and Paris VII, 4 Place Jussieu, 75252 Paris, France*

**Abstract**

We have put into evidence the existence of an antiferromagnetic coupling between iron epilayers separated by a ZnSe crystalline semiconductor. The effect has been observed for ZnSe spacers thinner than 40Å at room-temperature. The coupling constant increases linearly with temperature with a constant slope of ~5.5x $10^{-9}$ J/$m^2$K. The mechanisms that may explain such exchange interaction are discussed in the manuscript. It results that thermally-induced effective exchange coupling mediated by spin-dependent on and off resonant tunnelling of electrons via localized mid-gap defect states in the ZnSe spacer layer appears to be the most plausible mechanism to induce the antiferromagnetic coupling.

## 1. Introduction

The special interest in hybrid ferromagnet/semiconductor heterostructures arises from their plausible wide uses in spintronic devices. The active research on compatible materials to composed junctions and the more accurate growth conditions had caused that these new materials can reach technological standards for their application. Nowadays, one of the most difficult challenge in these kind of devices is to build clear and sharp junction interfaces. This desired property improves, mainly, the eletronic injection through the junction and it also allows a better understanding of other properties in these systems, in particular, the magnetic ones.

The magnetic coupling between two ferromagnetic layers across a non-magnetic spacer has been intensively investigated in the last years, both experimentally and theoretically. An oscillatory ferromagnetic/antiferromagnetic coupling has been observed in metallic multilayers[1]. The interlayer exchange coupling (IEC) cannot be simply explained by the Rudderman-Kittel-Kasuya-Yoshida (RKKY) theory as was initially proposed. The physical origin of the coupling effect has been attributed to quantum interferences due to spin-dependent reflections at the spacer boundaries[2-4]. The quantum well state (QWS) nature of the interlayer coupling in metallic systems was experimentally confirmed by magnetic measurements[5,6] and photoemission experiments[7]. The observation of an oscillatory behaviour in the exchange coupling as a function of ferromagnetic layer thickness[5,6], $t_F$, and the strong temperature dependence of the IEC strength support QWS models[8]. Nevertheless, P.Bruno[2] has shown that the quantum interference effects can be included in a modified RKKY model. He also proposed a unified treatment of the interlayer coupling through metallic and insulating spacer layers, by introducing the concept of a complex Fermi surface.

Whereas the Bruno's theory applies to crystalline systems, there are experimental finding of exchange coupling across non-metallic spacers in polycrystalline materials and even amorphous, namely a-Si[9], a-Ge[10], and a-ZnSe[11]. High-quality epitaxial semiconductor EuS/PbS/EuS trilayers exhibit antiferromagnetic coupling, but their strength decreases with temperature consistently with the power-law dependence of the EuS magnetization[12]. Other semiconductor epitaxial systems such as (Ga,Mn)As/(Al,Ga)As/(Ga,Mn)As[13] and (In,Mn)/InAs/(In,Mn)As[14] are ferromagnetic exchange coupled for rather long spacer thickness range, typically around 30 nm. Strong antiferromagnetic coupling, exponentially decaying with the thickness of Si spacer has been observed for Fe/Si/Fe epitaxial structures[15]. However, the high chemical reactivity at the Fe/Si interfaces always complicates the physics of the system. A concrete experimental evidence of room-temperature antiferromagnetic interlayer coupling by equilibrium quantum tunneling of spin-polarized electrons between the ferromagnetic layers has been reported on epitaxial MgO(100)/Fe/MgO/Fe/Co magnetic tunnel junctions[16].

In this paper, we present a detailed study of the interlayer magnetic coupling across a semiconductor barrier as a function of the spacer thickness and temperature. With this purpose, we have grown trilayers with bottom and top iron layers separated by a ZnSe epilayer with a continuosly variable thickness (wedge). Such wedge allows us to rule-out growth unavoidable deviations from one sample to another present in the case of samples with different thicknesses are prepared. We would like to emphasize that the Fe/ZnSe system can be classify as a rare successful example of ferromagnetic metal-semiconductor epitaxy where chemistry and magnetic properties of the Fe/ZnSe(001) interfaces remain stables up to 300 $^{\circ}$C[17].

## 2. Experimental details

The Fe/ZnSe/Fe samples have been prepared by Molecular Beam Epitaxy (MBE) in a multi-chamber growth system. First, a 3000Å thick undoped GaAs buffer layer was grown onto a GaAs(001) substrate with at the end, the stabilization of the β(2x4) As-rich reconstruction. The sample was then quickly transferred to the II-VI growth chamber where an also undoped 100Å ZnSe epilayer was deposited at 220$^{\circ}$C using alternate layer epitaxy [18]. The c(2x2) Zn-rich surface was stabilized at the end of the growth. Over this reconstructed surface the bottom iron layer was grown (65Å thick) at 180°C using a Knudsen cell with a grown rate of about 1.4 Å/min. This iron film is uniform and flat, as reported in previous studies performed by scanning tunnelling microscopy[19]. Next, the ZnSe wedge was grown at 220°C following the specific procedure already described elsewhere[20]. The wedge thickness was varied between 25Å and 80Å, with a slope of about 1~2 Å/mm and oriented along the ZnSe [110] direction. The thickness of the wedge have been also controlled by transmission electron microscopy (TEM). Finally, the top iron layer was deposited with a thickness of 140Å and substrate temperature of 180°C. An amorphous ZnSe cap layer deposited at room temperature was used to protect the samples against air exposure.

We have also studied other trilayers[21] with constant and homogeneous ZnSe thicknesses, between 30Å and 80Å, in order to check the reproducibility of our results obtained in wedge sample. Thin reference Fe layers were also grown in order to determine the magnetic parameters for single isolated layers. Reflection high-energy electron diffraction (RHEED), X-ray photoelectron spectroscopy (XPS) and scanning tunnelling microscopy (STM) in-situ facilities have been used to characterize the growth of the samples. These results are published elsewhere[18,20].

The Fe/ZnSe(wedge)/Fe trilayers were cleave into thin slices, perpendicular to the wedge direction. The ZnSe thickness gradient within each sample was always smaller than 3Å.

The magnetization measurements were carried out in a SQUID magnetometer. The field was applied parallel to the film plane along the characteristic crystallographic axis of the substrate. We have used the *minor hysteresis loops* method to evaluate the magnetic coupling between the Fe layers. In this method, once fixed the magnetization direction of the hard ferromagnetic layer, minor loops are measured sensing only the magnetization of the softer ferromagnetic layer. If there is a magnetic coupling between the ferromagnetic layers, the minor loop is displaced along the field axis. The net sign of the displacement with respect to the zero-coupling position, positive or negative, results in an evidence of a ferromagnetic (F) or antiferromagnetic (AF) coupling respectively. The shift modulus is known as the compensation field ($H_{comp}$) and corresponds to the magnetic field at which the magnetostatic energy of the softer magnetic layer and the coupling energy between layers are identical. The coupling constant $J$ is then quantitatively estimated as the energy difference, per unit area, between parallel and antiparallel magnetization alignment; i.e., $J = ½\ t_F.M_S.H_{comp}$ with $t_F$ being the softer layer thickness. We arbitrarily disregard the high order coupling constants in our analysis.

In our measurements, the magnetic field is determined by converting the current applied to the superconducting coil. As the compensation and coercive fields involved in the problem are comparable with the typical remanent fields ($H_{rem}$) of the superconducting coil, a systematic procedure was employed to minimize and to estimate $H_{rem}$.

The first experiments were performed following the *degauss shield* and *reset magnet* procedures so as to minimize $H_{rem}$ to around - 0.18 Oe, comparable to local earth magnetic field. This value was obtained using the *low-field flux gate* option. In the following, these procedures were not used because the remanence measured after cycling the magnetic field between 1.5 kOe → 0 Oe and –1.5 kOe → 0 Oe was systematically higher. The saturation field of the iron layers was always lower than 1.5kOe and the mean $H_{rem}$ results 1.5Oe ± 0.25Oe. A small asymmetry between positive and negative $H_{rem}$ was found but this difference lays between the field error bars. The mean $H_{rem}$ value was used to correct the current-converted measured magnetic field in the low field region.

Ferromagnetic resonance experiments have been also performed at Q-band (~33 GHz) in a Bruker spectrometer. The angular dependence of the resonance spectra has been studied in the in-plane ([100] to [010]) geometry.

3. Results

We will first describe the structure and interface morphology of the Fe/ZnSe-wedge/Fe epilayers. Next, the magnetic properties of the Fe/ZnSe/Fe trilayer will be presented. The existence of a magnetic coupling strength across the ZnSe spacer, as well as its dependence with the barrier thickness and temperature will be shown.

3.1 Structural characterization

The first Fe layer grows epitaxially on top of the pseudomorph ZnSe(001) thin epilayer grown on GaAs(001) substrate. According to RHEED and XPS analysis, the 65Å thick Fe(001) films are completely relaxed and display high quality and uniformity[22]. The STM images display a surface morphology of Fe epilayers with very small roughness even at micrometric scale. This picture remains almost unchanged for iron films from 10 to 100Å thick 100Å. Next, we have grown the crystalline ZnSe(001) epilayer with a wedge-shape on top of this first Fe layer. The procedure used to prepare the ZnSe wedge layer has been already used previously to investigate the ZnSe growth on GaAs(001) and it is published elsewhere[20]. monitored by in situ RHEED experiments. The minimum thickness of the ZnSe-wedge layer was set to 25Å in order to avoid pinholes and ex-situ TEM analysis has been used to precisely calibrate the thickness and consequently the gradient of the ZnSe-wedge spacer. The ZnSe thickness shows a linear behaviour with position along the wedge. According to TEM micrographs the thickness gradient has a slope of about 1 ~ 2 Å/mm[20].

Fig. 1 shows cross-sectional TEM images of samples with $t_{ZnSe}$= 40Å and $t_{ZnSe}$= 80Å where abrupt and atomically flat Fe/ZnSe and reverse ZnSe/Fe interfaces can be clearly observed. In both cases, the ZnSe spacer layer is rather uniform and shows no evidence of pinholes or disruptive defects. The top Fe layers exhibits a poor crystalline quality comparatively to the bottom one with the structural disorder and defects homogeneously distributed along the iron epilayer. Similar results were found for the Fe(140Å)/ZnSe($t_{ZnSe}$)/Fe(65Å) structure grown under similar conditions.

**3.2 Magnetization measurements**

In Fig. 2(a) the magnetization versus applied field curves for three samples with $t_{ZnSe}$ = 25Å, 37Å and 45Å are shown. The magnetization loops show two-steps in all the samples, corresponding to the magnetization reversal in the top and the bottom iron layers. An evidence that the two Fe layers are weakly coupled is the observation of the magnetization plateau slightly higher than the expected value $M/M_S$ = 0.37; i.e., the ratio between difference of the saturation magnetization of bottom and top Fe layers, and the sum of the saturation magnetization of the Fe layers.

In Fig. 2(b) the minor hysteresis loops superimposed to the full hysteresis for the $t_{ZnSe}$ = 37Å sample is shown. A slight shift of the minor loops to negative $H_{comp}$, i.e., a weak antiferromagnetic coupling between Fe layers, is observed at room temperature.

The tiny coupling is observed for all the studied samples with the ZnSe thickness varying from 25Å to 80Å.

The magnetic coupling between Fe layers have been quantitatively estimated from $H_{comp}$. The coupling strength constant ($J$) as a function of the ZnSe thickness is shown in Fig. 3. The coupling is stronger for thinner spacer layers. However, it is difficult to deduce from the data a functional dependence of the coupling strength with $t_{ZnSe}$. To better explore this particular point, we have used FMR experiments that have allowed us to obtain a much clearer picture of this behaviour (see below).

The temperature dependence of the magnetic coupling for $t_{ZnSe}$ = 40Å and 80Å, obtained from the SQUID measurements, is depicted in the Fig. 4. As can be seen, the coupling strength increases linearly with temperature and the slope is approximately the same for the two samples. The discussion of this effect will be performed in the next section.

The temperature dependence of the IEC across metallic and non-metallic spacer layers is currently ascribed to two mechanisms: i) thermal excitations of electron-hole pairs across Fermi level as described by the Fermi-Dirac function, and ii) thermal excitation of spin waves in the magnetic layers and particularly at their interfaces. Specifically, for ZnSe semiconducting spacer layer the former mechanism is extended to include: a) a spin-dependent thermal repopulation of levels close to the Fermi surface via overlapping across the spacer layer of localized, weakly bound electron states situated at or near the two interfaces[23] and b) spin-dependent thermal population of impurity/defect mid-gap states via resonant and non-resonant tunneling[24].

### 3.3 Ferromagnetic resonance measurements

The FMR spectra have been measured with the static magnetic field applied in the plane of the samples. The samples measured by FMR belong to the same wedged-sample examined by SQUID. The spectra are composed of two resonance modes (Fig. 5), whose intensities are strongly affected by the magnetic field direction and the ZnSe spacer thickness when the Fe layers are magnetic coupled as we will see in Subsection 3.3.2.
In order to analyze the measured data we propose a general expression for the free energy density $F$ of a coupled trilayer:

$$F = t_1 F_1 + t_2 F_2 - J M_1 . M_2 \qquad (1)$$

where $F_i$ and $t_i$ denote the free energy density and the nominal thickness of each magnetic layer, respectively. The last term corresponds to the bilinear magnetic exchange between the layers. $M_i$ is the saturation magnetization of the $i$ iron layer, being $M_1$ equal to $M_2$ (1714 emu/cm$^3$).

In order to deduce the free parameters of Eq. (1), we have studied the angular dependence of the resonance fields as well as the relative intensities of both modes.

### 3.3.1 FMR parameters of the Fe layers

We have calculated the $F_i$ parameters by fitting the angular dependence of the resonance field discarding the influence of the exchange coupling. As we will show below, the interaction between iron layers is so weak that it does not affect the resonance fields.

The in-plane (IP) angular dependence of the resonance fields of the trilayer with $t_{ZnSe}$ = 45 Å is shown in Fig. 6(a). The samples with $t_{ZnSe}$ = 48 Å and 50Å exhibit the same behaviour. The plot shows that in both iron layers, the main contribution to the anisotropy of the system is a four-fold anisotropy term associated with the magneto-crystalline anisotropy of the bcc Fe structure. The mode with smaller resonance field corresponds to the thinner (bottom) layer and the other one (larger resonance field) to the thicker (top) one. This conclusion is obtained from the calculation of the lines intensity, which is proportional to the magnetization of the layers[25]. The ratio between the intensities of both modes varies from 0.42 to 0.44 along the samples series, in good agreement with the nominal ratio of the iron layer thickness, 0.46. This conclusion can be also deduced from the position of the modes. In fact, three contributions can shift the resonance fields: the layers magnetization, the demagnetizing field and a uniaxial out-of-plane contribution which comes from the surface[26]. The first two terms give the same contribution for the two layers, i.e. both layers have the same magnetization and geometry, but the last term becomes important in thin films due to the higher relative contribution of the films surface over its volume. In our case, we have taken into account this result for modelling the resonance line. We confirm that the out-of-plane anisotropy is more important for the thinner layer, as expected, and therefore the resonance line of this layer is shifted upward.

A small in-plane uniaxial anisotropy has been measured in the [110] direction only for the thinner iron layer. This contribution is induced at the ZnSe/Fe interface and has been previously reported for ZnSe/Fe and GaAs/Fe structures[27,28]. This anisotropy has not been observed in the iron thicker layer, probably, due to the fact that the anisotropy scales with $1/t$ and thus it is negligible for the top layer.

In Fig. 6(b), we display the angular dependence of the resonance field of the sample with $t_{ZnSe}$=25Å. A similar behaviour has been observed in the sample with $t_{ZnSe}$=31Å. We can see that the angular dependence of the thicker layer changes drastically respect to 7(a). This issue will be further discuss below.

To take into account the symmetries of our results we propose the following free energy density for the iron layers:

$$F_i = -\mu_0 \mathbf{H}.\mathbf{M}_i + \frac{1}{4} K_{4i} \left[ \sin^2(2\Theta) + \sin^4 \Theta \ \sin^2(2\phi) \right] + \frac{1}{2} M_{eff,i}^2 \cos^2 \Theta \\ + K_{u,i} \cos^2(\phi - \phi_{u,i}) \sin^2 \Theta \quad (2)$$

where the first term is the Zeeman interaction of the iron moments with the external magnetic field, the second term describes the four-fold anisotropy, the third one is the shape anisotropy corrected by an uniaxial anisotropy, $K_n$, that favours an out-of-plane (OOP) orientation of the magnetization. The last term corresponds to the in-plane uniaxial anisotropy $K_u$. $\Theta$ is the polar angle, between $M$ and the normal to the film plane, $\phi$ is the azimuthal angle between $M$ and [100] and $\phi_u$ is the angle between the uniaxial in-plane direction and [100]. $M_{eff}$ is the effective magnetization, given by the relation $\frac{1}{2} M_{eff}^2 = \frac{1}{2} M^2 + K_n$.

The equilibrium angles of $M$ are obtained from the minimization of $F$. The resonance frequency is given by[29]:

$$\left(\frac{\omega}{\gamma}\right)^2 = \frac{1}{M^2 \sin^2 \Theta}[F_{\Theta\Theta}F_{\varphi\varphi} - F_{\Theta\varphi}^2] \quad (3)$$

where the subscripts indicate partial derivatives, evaluated at the equilibrium angles $\Theta$ and $\varphi$; $\gamma = g \cdot \mu_B / \eta$ is the gyromagnetic ratio and $\omega = 2\pi\nu$.

In Fig. 7, we show the evolution of the different parameters of the free energy density with the ZnSe thickness. The parameters for the trilayers with spacer thickness larger than $t_{ZnSe}$ = 40Å remains almost constant, in agreement with previous works[26,30]. As we remark above, $M_{eff}$ and $K_4$ of the thicker iron layer change notably for samples with thinner spacers. These two terms have surface contribution[31], and so they are very sensitive to changes in interface morphology and magnetism. In particular, in thinner-spacer samples the surface roughness of the thick Fe layer is expected to be enhanced, because a few monolayers of ZnSe are not enough to smooth the strain effects at the ZnSe/Fe$_b$ interface. Note that this effect is also seen in the magnetic measurements through the change of the coercive field.

### 3.3.2 Interlayer exchange coupling

The coupling[25] between the magnetic layers may change the position of the resonance field, $Hr$, the line-width, $\Delta H$, and the intensity of the FMR signal, $I$. Unfortunately, in our case, the coupling seems to be too small to induce a significant change of the line width or a shift in the resonance field line position.

For magnetic couplings smaller than $1\times 10^{-4}$ J/m$^2$, the shift of the resonance mode is of the order of 10Oe. The broad lines and the measurement noise difficult the detection at these very small resonance field variations. Also, both resonance modes are very close for samples with $t_{ZnSe}$ = 31Å and 25Å and the lines cross each other when the azimuthal angle is changed. When the lines are crossing each other in these samples, it is expected that the changes of the line positions are larger (about ~150G for $J=1\times 10^{-4}$ J/m$^2$). However, no shift is noticed because the lines are too large and hide any change of the resonance fields ($\Delta H$ ~140G and $\Delta H$ ~200G for the thinner and thicker layers, respectively).

The resonance line-width is mainly extrinsic and it is attributed to a distribution of magnetic anisotropies[32]. We have not observed any systematic change of the line width with the spacer thickness.

The intensity of the FMR signal is proportional to the magnetization of the sample and it is defined as follow[25]:

$$I = M \frac{\left[\int_0^{tb} m_b(z)\,dz_1 + \int_0^{tt} m_t(z)\,dz_2\right]^2}{\int_0^{tb} m_b^2(z)\,dz_1 + \int_0^{tt} m_t^2(z)\,dz_2} \quad (4)$$

where $m(z)$ denotes the amplitude of the time-dependent magnetization component and it is integrated in the $z$ direction (normal to the samples' plane). $M$ is assumed to be homogeneous along the layers. In spherical coordinates, $m_i = m_{i,\theta}\hat{\theta} + m_{i,\phi}\hat{\phi}$, where $i=b,t$ denotes the bottom and the top iron layer, respectively. The thickness of both layers is indicated by $t_b$ and $t_t$. The intensity of each mode is obtained from Eq. (4). In the case of uncoupled systems, the intensity of each line is proportional to the magnetization of each individual layer. On the other hand, when the system is coupled, the intensity of the modes varies, depending on the strength of the coupling. In this case, the modes entangle the ions' precession of both layers, being commonly named acoustic and optical ones[25]. The acoustic (optical) mode is associated to the low (high)-energy one. When an antiferromagnetic (AF) coupling is present the low energy mode corresponds to the out-of-phase precession of the two magnetic moments. The out-of-phase (OP) precession favors an antiparallel alignment of the time-dependent component of the magnetic moments. Thus, for the AF coupling the out-of-phase mode has a lower energy respect to the in-phase precession. In the ferromagnetic coupling case, the energetic position of the two modes is inversed.

The intensity of the modes depends on the contributions of the precessing moments. Therefore, the intensity of the out-of-phase mode will be lower than the in-phase mode because the total time-dependent magnetization is lower for an antiparallel alignment of the magnetic moment of the layers. When the layers are identical, the magnetization of one layer compensate the magnetization of the other one and so the intensity of the out-of-phase mode is zero.

When the coupling is very weak, as it is in our system, the mode lines preserve their uncoupled characteristics and one can associate each mode to a particular iron layer. The modes have essentially an independent layer-like behaviour. Then, their intensities vary slightly respect to the uncoupled situation.

From the analysis of the ratio between mode intensities we have deduced the coupling strength through the ZnSe barrier. The intensity of the different modes is given by

$$I_j = \frac{2M(t_1 m_{\varphi 1} + t_2 m_{\varphi 2})^2}{t_1(m_{\varphi 1}^2 + m_{\theta 1}^2) + t_2(m_{\varphi 2}^2 + m_{\theta 2}^2)} \qquad (5)$$

where the sub-index $j$ labels the mode related to the top and bottom layer, respectively. To derive the last equation from Eq (1), we assume that $m$ is homogeneous in the whole layer. The time dependent component of the magnetization, $m$, is calculated from a generalized Smit-Beljers equation taken from Ref. 33.

We compute the ratio of the thinner layer line intensity, $I_b$ (in the uncoupled case), to the thicker one, $I_t$. This ratio varies, depending of the strength of the coupling as well as the resonance field of the modes.

In Fig. 4, we show the spectra of the sample with $t_{ZnSe}$=25Å with the magnetic field applied in the [100] and the [110] directions. We clearly see how the bottom layer mode intensity, $I_b$, changes with the field direction. It can be noticed that $I_b$ decreases when the resonance field of the mode is larger than that of the thicker layer mode. This situation corresponds to the magnetic field applied in the [110] direction.

In Fig. 8 the calculated $I_b/I_t$ ratio as a function of the coupling constant strength and sign is plotted. From this figure we can appreciate that in the case of ferromagnetic coupling, the ratio $I_b/I_t$ for the coupled system is smaller than that of the uncoupled one when $H$ is oriented along [100], while it is larger for [110] direction. The opposite situation is found for antiferromagnetic coupled structures and matches our experimental results. The explanation of the intensity dependence on the coupling sign and field directions is the following: in samples with $t_{ZnSe}$= 25Å and 31Å, the resonance fields of both layers alternate their relative positions when $\phi_H$ is varied as can be seen in Fig. 6(b). In the [100] direction, the resonance field of the thinner layer is lower than the one measured for the thicker layer while for the [110] direction, the opposite situation is observed. For thicker barrier thickness, the position of both modes remains always parallel. In Fig. 8 the coupling constant calculated from the measurements, for the samples with $t_{ZnSe}$= 25 and 31 Å are indicated.

The coupling constant, deduced from FMR experiments is plotted in Fig. 2. FMR results indicate that the coupling is negligible for ZnSe spacers thicker than 40 Å at room temperature. A small antiferromagnetic coupling is measured for samples with thinner spacers but rapidly decays as the ZnSe thickness is increased. The coupling constant strength is of the order of $10^{-5} J/m^2$ and about one order of magnitude smaller than the one measured in metallic structures[33].

## 4. Discussion

Our results put in evidence the existence of a temperature-dependent antiferromagnetic coupling between iron layers, across a single-crystalline ZnSe semiconducting barrier. Extrinsic mechanisms as pinholes and dipolar fields may induce a magnetic coupling across non-magnetic spacers. A direct coupling through pinholes has been discarded due to the antiferromagnetic nature of the measured interaction.

In the thin-film geometry, the magnitude of the magnetostatic coupling for ideal and smooth layers is vanishing small. However, the magnetostatic coupling becomes significant if a surface roughness is considered. In this case, steps, ripples and other features of the magnetic layers surface result in stray magnetic field lines that couple the neighbouring layers; i.e., this is the so-called Orange-Peel (OP) or Néel coupling. This coupling leads to an effective ferromagnetic or antiferromagnetic field depending on the roughness topology. We have calculated the Orange-Péel coupling strength for the Fe/ZnSe/Fe trilayers. TEM and STM images allow us to estimate the roughness amplitude at the different interfaces ZnSe/$Fe_b$, $Fe_b$/ZnSe barrier, ZnSe barrier /$Fe_t$, $Fe_t$/cover ZnSe of the structure. The calculation has been performed for the worst situation, that is assuming one monolayer roughness at the $Fe_b$/ZnSe barrier and the $Fe_t$/ZnSe barrier interfaces, with a wavelength of 50nm. This amplitude is increased to 1nm for the outer Fe interface of the top layer. The ZnSe/$Fe_b$ interface is very flat and no roughness at this interface has been considered in the calculation of the coupling. Using the formula given in Ref. 34 and the roughness parameters mentioned above, the OP coupling strength results to be -14μJ/m$^2$. The coupling is of the same order of magnitude of the measured one, but ferromagnetic and therefore does not match with our results.

Moreover and independently of its exact formula, this effective dipolar coupling is proportional to the magnetization of the magnetic layer, which decreases with increasing temperature. Thus, the OP dipolar coupling is expected to decrease with increasing temperature and so, it cannot explain the temperature dependence of coupling measured in our samples. A variation of less than 2% in the OP coupling is expected for the temperature range of our measurements.

Therefore, we have explored theoretical models based on intrinsic coupling mechanisms to analyse our results. P. Bruno proposed in Ref. 2, an extension of his former RKKY interlayer coupling model for metallic and insulating barriers. In spite of the unified treatment, the thickness and temperature dependence of the coupling is drastically different in both cases. The coupling dependence with spacer thickness is oscillatory in metallic structures while it decreases exponentially in the case of an insulator barrier. P. Bruno deduced the existence of an antiferromagnetic coupling for spacer thickness $t > 1$nm. The coupling strength decays rapidly from 10μJ/m$^2$ to zero, as the barrier thickness is increased from 1.2nm to 2.3nm. Bruno's calculations were performed for relatively low barriers ($U - \varepsilon_F = 0.1$ eV) and assuming that the electrons that mediate the coupling have a s-character. Similar calculations assuming theoretical parameters for Fe/MgO/Fe system ($U - \varepsilon_F = 1$ eV) were performed by Faure-Vincent *et al.* in Ref. 16. The barrier height of our samples is higher than the value used for the Bruno´s calculations, but quite close to the value used in Ref. 16. According to our previous photoemission experiments[35], the Fe-Fermi level position is stabilized at 1.6 eV above the valence-band maximum of undoped ZnSe, i.e., a barrier height of $U - \varepsilon_F = 1.1$ eV. Our results agree qualitatively with the spacer thickness dependence of the coupling strength predicted by this model i.e.: we have found an exponential decrease of the antiferromagnetic coupling strength with the ZnSe spacer thickness.

The temperature dependence of the antiferromagnetic coupling depicted in the Figure 4, does not seem to show the behavior reported in Ref. 2. The exchange coupling measured for our samples increases linearly with increasing temperature. We want to remark, however, that the larger variation of the coupling strength reported by P. Bruno, is observed above room temperature and we were not able to measure the coupling in this temperature range.

Landolt and coworkers[11] measured a thermally-induced antiferromagnetic exchange coupling across amorphous ZnSe in the spacer thickness range between 18Å and 25Å. The coupling changes to ferromagnetic below 15Å and above 30Å at any investigated temperature. The magnitude of the antiferromagnetic exchange coupling strength is very small (< 18 $\mu J/m^2$) exhibiting a thermal saturation above ~100 K. In our samples with crystalline ZnSe spacers, the magnitude of the antiferromagnetic coupling strength is much larger at room temperature and presents no evidence of thermal saturation. We have studied samples with ZnSe spacer thickness range above the thickness region where thermally-induced antiferromagnetic exchange coupling was found in a-ZnSe spacer to minimize the effect of extrinsic factors such as layers thickness fluctuations due to interface waveness, crystal/interface quality and pinholes. The thickness waveness and pinholes formation seem to be more critical in our epitaxial structures than in samples with amorphous spacers. Landoldt *et al* analyse their data in terms of large molecular orbitals built across the spacer layer[23]. This explanation is not suitable for our case because in our samples the defect/impurity density is lower and the spacer thickness range is much larger comparatively to those found for a-ZnSe.

We have also analysed the double-quantum-well proposed by Hu and coworkers[36] to explain the exchange coupling in trilayered structures. These authors report the existence of two kind of interlayer couplings, a resonant and a non-resonant one. The first is, alternatively, ferromagnetic and antiferrromagnetic changing without a definite periodicity but depending on the ferromagnetic layer thickness. The coupling strength varies from $-1 \times 10^{-3} J/m^2$ to $2.5 \times 10^{-3} J/m^2$ as the iron layer thickness varies from 0nm to 20nm, keeping the barrier thickness constant at 1nm. The non-resonant *J* is much smaller than the resonant one (less than 1$\mu J/m^2$), and varies drastically with the iron thickness. The calculated coupling strength for both cases is different from the measured in our samples. Moreover, the resonant case is observed for very thin magnetic layers which is not our case. The variation of the calculated *J* with temperature is strongly dependent on the tunnelling mechanism, the height of the barriers and the thickness of the magnetic layers. The authors explain the temperature dependence of the antiferromagnetic coupling in terms of spin-flip non-resonant tunnelling.

**5. Conclusion**

We have presented experimental evidence of magnetic coupling through a semiconducting barrier in the Fe/ZnSe/Fe system. The samples were epitaxially grown by MBE and display very small surface roughness. A weak antiferromagnetic coupling between the iron layers across the ZnSe barrier was measured through magnetization loops and FMR experiments. The coupling strength decreases rapidly as the $t_{ZnSe}$ is increased, becoming negligible beyond $t_{ZnSe}=45$Å at room temperature. The Orange-Péel dipolar coupling calculated for the samples is always ferromagnetic and thus ruled out as the origin of our results. Therefore, we explained our results in terms of an intrinsic coupling between ferromagnetic layers across a non-metallic barrier by spin-polarized quantum tunneling of electrons. The thickness dependence of the coupling agrees with theoretical models proposed for these structures[2].

As a particular challenge, we propose the tailoring of the epitaxial growth of this system and other ferromagnetic/semiconductor/ferromagnetic systems to investigate the exchange coupling in a wide spacer thickness range and the application of a bias voltage to control thermal excitations and population of the electronic states.

**Acknowledgments**

The authors thank partial financial support from bilateral program Capes-Cofecub, ACI-Nanosciences, CNPq, FAPESP, ANPCYT (PICT N$^0$ 3-6340), CONICET (PIP N$^0$ 2626) and Fundación Antorchas. J.M. acknowledges to A. Butera for scientific assistance. J.M. and L.B.S. acknowledge to J. Pérez and R. Benavídez for technical assistance. L.B.S is a member of CONICET, Argentina.

**Figure Captions**

**Figure 1** - Cross-sectional transmission electron microcopy image of Fe(140Å)/ZnSe ($t_{ZnSe}$) /Fe (65Å) structures for $t_{ZnSe}$ 40Å (a) and 80Å (b). The arrows indicates the 100Å vertical scale.

**Figura 2** – (a) Magnetization loops for samples Fe(140Å) / ZnSe ($t_{ZnSe}$) / Fe(65Å) / ZnSe / GaAs(001) with $t_{ZnSe}$ = 25Å (open squares), 37Å (open circles), and 45Å (open triangles). (b) Superimposed to major loop of the sample with $t_{ZnSe}$ = 37Å is shown the minor loops after magnetization saturation at positive (circles) and negative (diamonds) fields. The magnetic field is applied parallel to [100] direction and the loops were performed at room temperature.

**Figure 3** – Magnetic coupling strength as a function of the ZnSe spacer layer according to magnetization (SQUID) and ferromagnetic resonance (FMR) measurements, performed at room temperature. Negative *J* values indicate antiferromagnetic coupling. The error bars for SQUID measurements are the point size.

**Figure 4** – Temperature dependence of the antiferromagnetic coupling of the samples with $t_{ZnSe}$ = 40Å and 80Å. Both samples have similar slopes of ~5.5x $10^{-9}$ J/m$^2$K.

**Figure 5** – FMR spectra of the trilayer sample with $t_{ZnSe}$= 25Å, when the static field is applied along the [110] and [100] direction. The spectra were normalized so as to have the same $I_t$ intensity.

**Figure 6** - Resonance fields as a function of the azimuthal angle of the external field. (a) for $t_{ZnSe}$=45Å, (b) for $t_{ZnSe}$=25Å. Open symbol are the resonance fields associated with the thicker layer and filled symbols with the thinner one. The data of the thinner layer dissapear for some angles beacause the two resonance lines merged in a only one for these directions.

**Figure 7 -** Magnetocrystalline ($K_4$) and out-of-plane ($K_N$) anisotropies as a function of ZnSe layer thickness.

**Figure 8 -** Calculated $I_b/I_t$ as a function of the coupling strength for both ferromagnetic and antiferromagnetic coupling. The vertical lines show the measured ratios for $t_{ZnSe}$=25Å and 31Å samples. The horizontal line shows the intensity ratio expected for an uncoupled system.

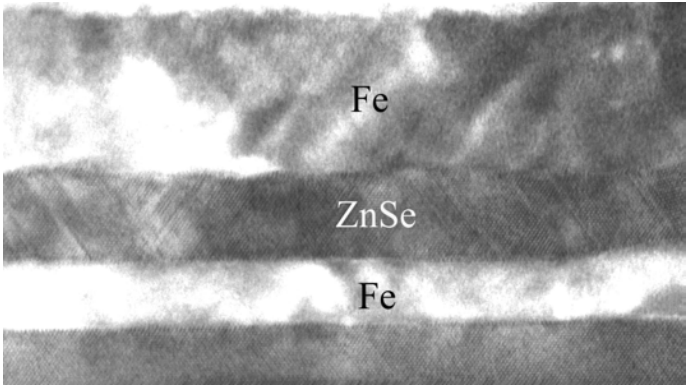

Varalda *et al.* Figure 1

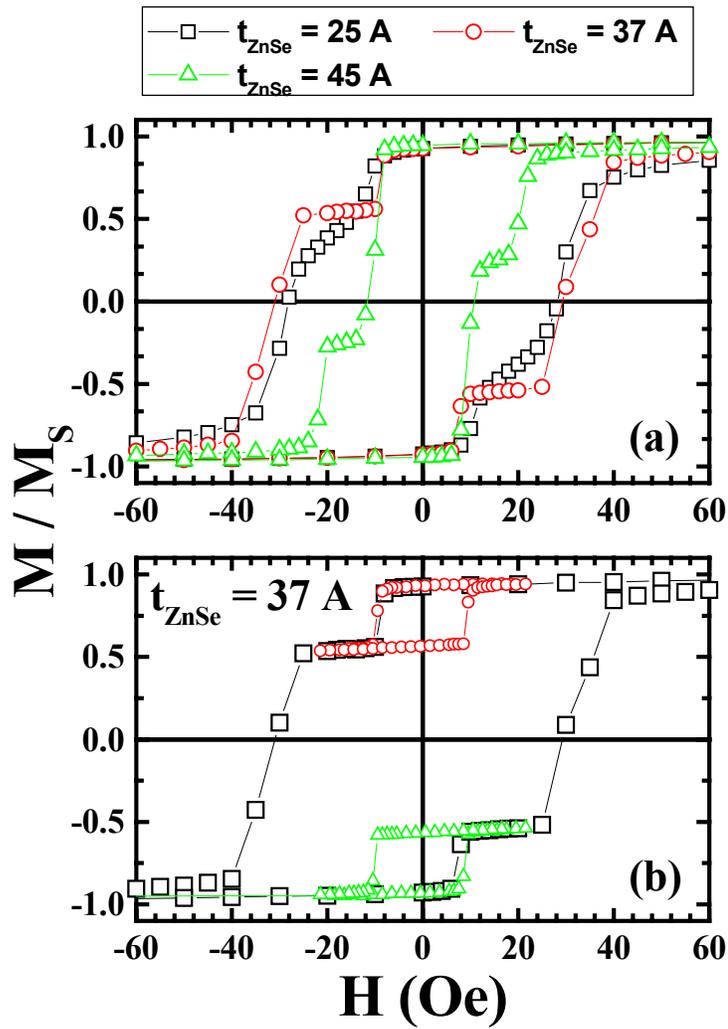

Varalda *et al*. Figure 2

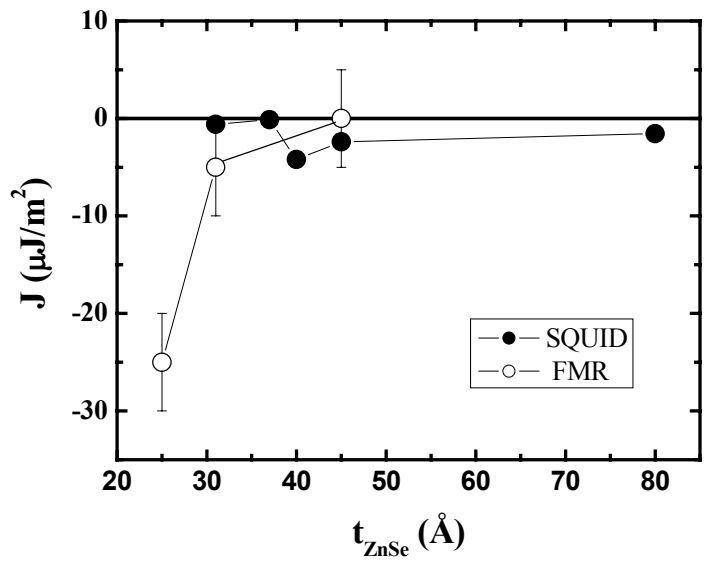

Varalda *et al.* Figure 3

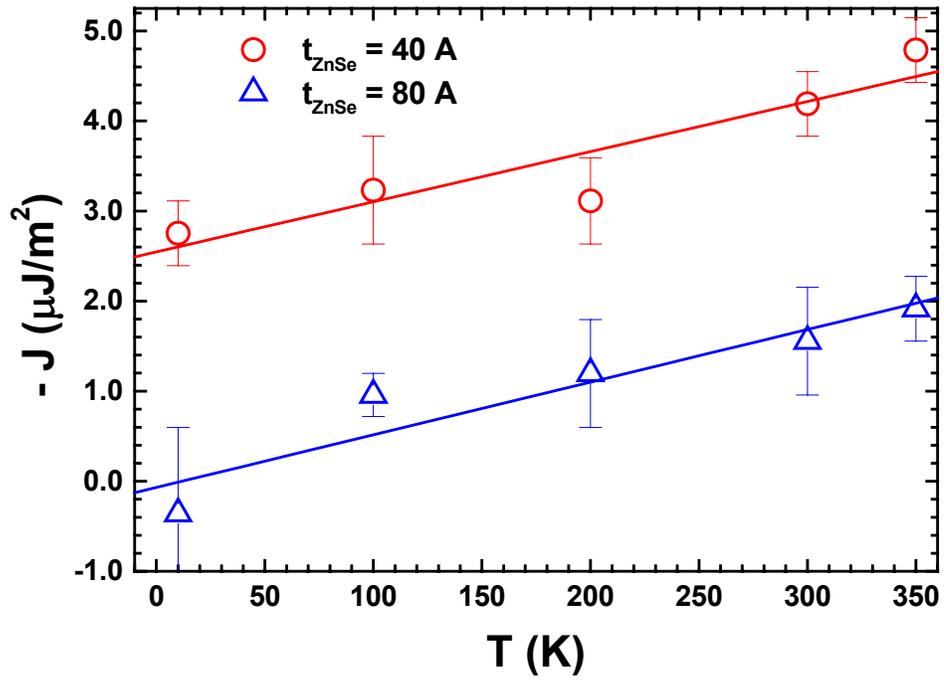

Varalda *et al.* Figure 4

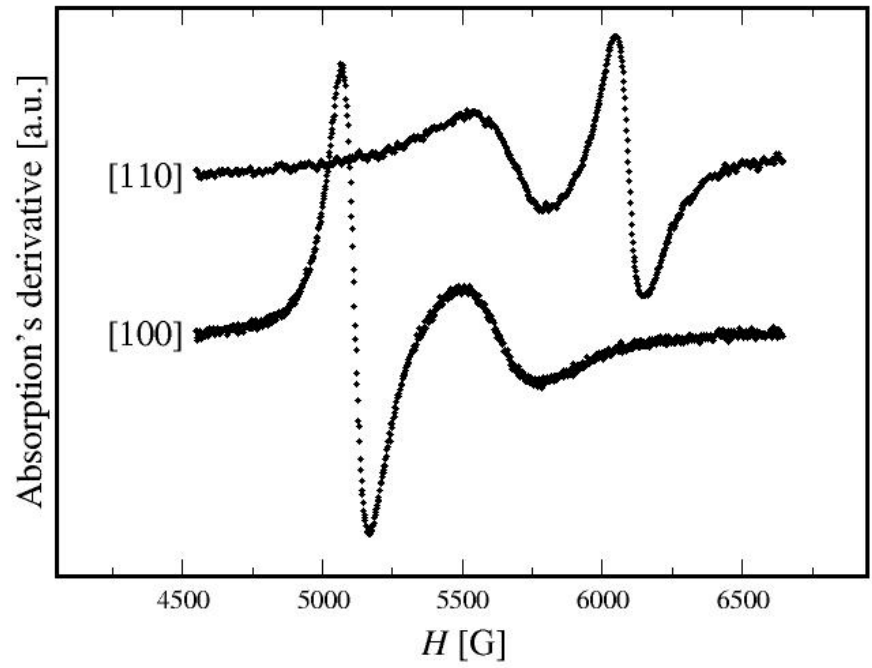

Varalda *et al.* Figure 5

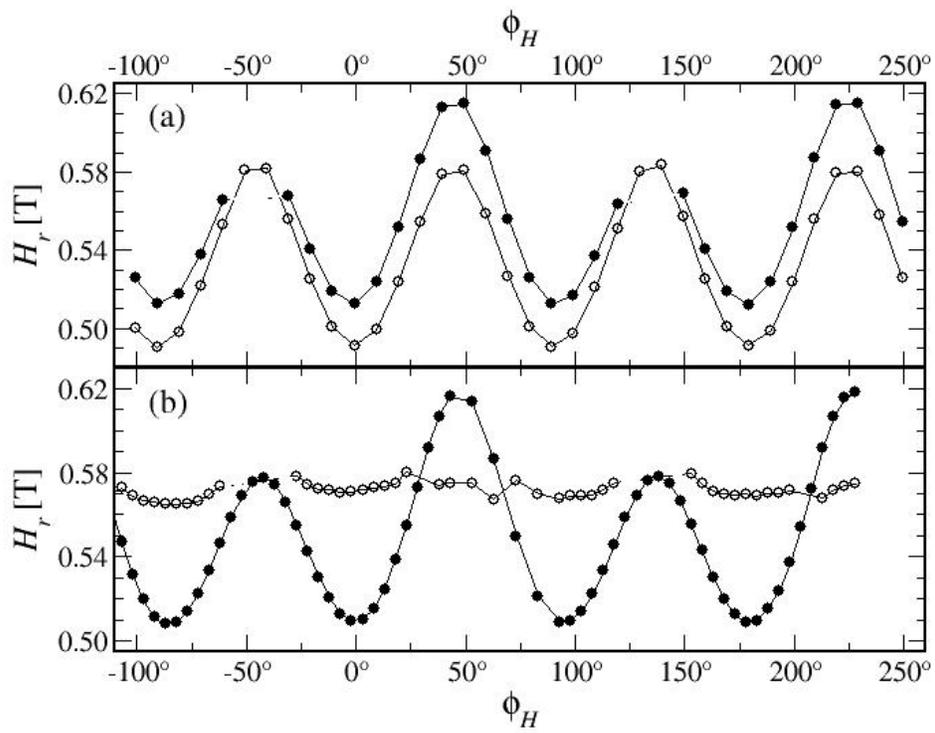

Varalda *et al.* Figure 6

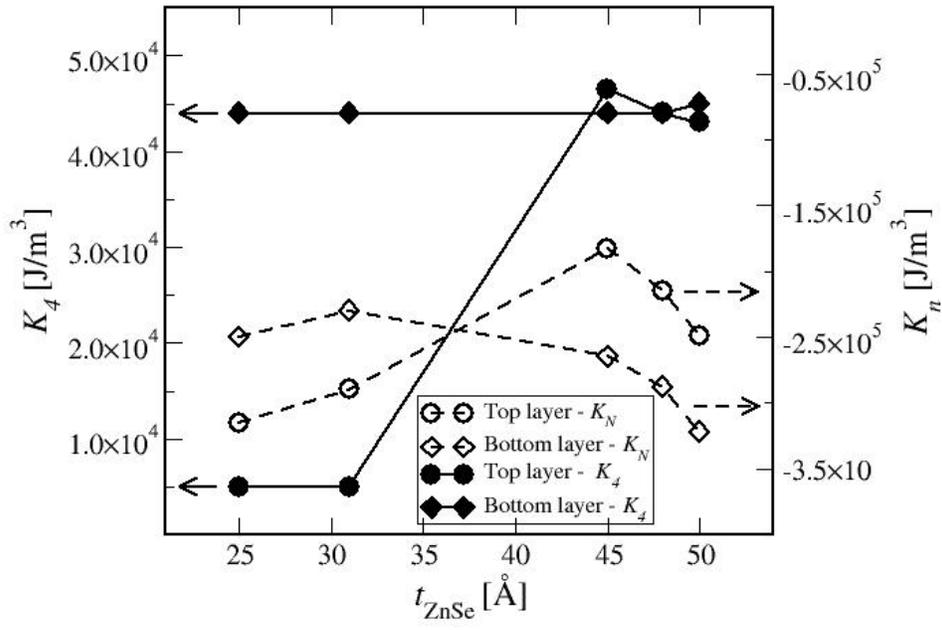

Varalda *et al.* Figure 7

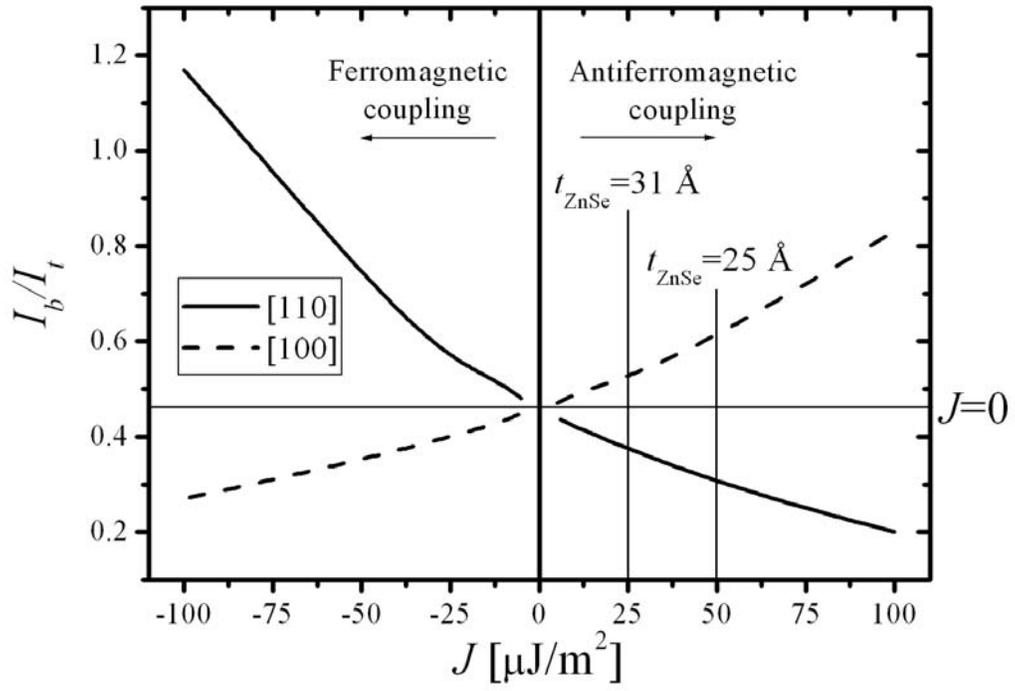

Varalda *et al*. Figure 9